\documentclass[twocolumn, journal]{IEEEtran}

\IEEEoverridecommandlockouts

\usepackage{color}
\usepackage{graphicx}
\usepackage{amsmath}
\usepackage{amssymb}
\usepackage{algorithm}
\usepackage{algorithmic}
\usepackage{amsmath}
\usepackage{multirow}
\usepackage{booktabs}
\usepackage{array}
\usepackage{amsthm}
\usepackage{stfloats}
\usepackage{caption}
\usepackage{subfigure}
\usepackage{bm}
\usepackage{setspace}

\usepackage{flushend}

\newcommand{\be}{\begin{equation}}
\newcommand{\ee}{\end{equation}}
\newcommand{\bea}{\begin{eqnarray}}
\newcommand{\eea}{\end{eqnarray}}
\newcommand{\ba}{\begin{array}}
\newcommand{\ea}{\end{array}}

\allowdisplaybreaks[4]

\title{End-to-End Learning for Symbol-Level Precoding and Detection with Adaptive Modulation
\thanks{R. Liu, Z. Bo, and M. Li are with the School of Information and Communication Engineering, Dalian University of Technology, Dalian 116024, China (e-mail: liurang@mail.dlut.edu.cn; zhubo@mail.dlut.edu.cn; mli@dlut.edu.cn).}
\thanks{Q. Liu is with the School of Computer Science and Technology, Dalian University of Technology, Dalian 116024, China (e-mail: qianliu@dlut.edu.cn).}
}

\author{Rang Liu,~\IEEEmembership{Graduate Student Member,~IEEE,}
        Zhu Bo,
        Ming Li,~\IEEEmembership{Senior Member,~IEEE,}\\
        and Qian Liu,~\IEEEmembership{Member,~IEEE}
        }

\begin{document}
\maketitle
\thispagestyle{empty}
\begin{abstract}
Conventional symbol-level precoding (SLP) designs assume fixed modulations and detection rules at the receivers for simplifying the transmit precoding optimizations, which greatly limits the flexibility of SLP and the communication quality-of-service (QoS).
To overcome the performance bottleneck of these approaches, in this letter we propose an end-to-end learning based approach to jointly optimize the modulation orders, the transmit precoding and the receive detection for an SLP communication system.
A neural network composed of the modulation order prediction (MOP-NN) module and the symbol-level precoding and detection (SLPD-NN) module is developed to solve this mathematically intractable problem.
Simulations verify the notable performance improvement brought by the proposed end-to-end learning approach.
\end{abstract}


\section{Introduction}
\vspace{0.2 cm}

Transmit beamforming/precoding design is very crucial to improve system performance by exploiting the spatial degrees of freedom (DoFs).
While most conventional block-level precoding (BLP) approaches attempt to suppress multi-user interference (MUI) based on the obtained channel state information (CSI), advanced symbol-level precoding (SLP) technology can convert harmful MUI into useful signals by utilizing both CSI and symbol information, thus significantly improving the quality-of-service (QoS) of multiuser transmissions \cite{Masouros TSP 2015}-\cite{Li ICST 2020}.
Although benefiting from both spatial and symbol-level DoFs, SLP approaches exhibit significantly higher computational complexities than their BLP counterparts.
In specific, the rationale of SLP is to design non-linear precoding for each transmitted symbol vector, which results in a huge number of precoders to be optimized during each channel coherent time.
To reduce the computational complexity, the authors in \cite{AL ITWC 2018} developed an efficient algorithm to simplify the optimization of each precoder with closed-form solutions.
Then, the authors in \cite{Xiao TCOM 2022} proposed a novel low-complexity grouped SLP approach, which reduces the number of required optimizations during each channel coherent time.
However, these approaches still struggle to support practical implementations that require very low complexity.

Concurrently, machine learning and specifically deep learning (DL) \cite{Shea TCCN 2017}-\cite{Zhang ICST 2019} have been applied to the physical layer of communication systems.
DL can effectively uncover non-linear correlations from sufficient numbers of data without resorting to mathematically tractable models, by which the computational complexity is transferred to the offline training process.
In light of these advantages, researchers have devoted to DL-based SLP designs in recent years \cite{Lei CL 2021}-\cite{Sohrabi ICASSP 2020}.
The authors in \cite{Lei CL 2021} developed a deep neural network (DNN) architecture for SLP designs of multi-user multi-input single-output (MU-MISO) systems, where the minimum communication QoS is maximized for a given transmit power.
A convolutional neural network (CNN) architecture was proposed in \cite{Zhu TVT 2021} to achieve the same goal.
For the power minimization problem in such systems, the authors in \cite{Mohammad GLOBECOM 2021} investigated to unfold the interior point method (IPM) iterative algorithm by utilizing a DL architecture.
Despite the fact that these approaches have much lower complexities than the corresponding optimization algorithms, all of them are developed based on fixed modulation and detection rules known to both the base station (BS) and the receivers.
As a result, the flexibility of SLP is significantly restricted and the resulting performance degrades.
To fully exploit the available DoFs of SLP, the authors in \cite{Sohrabi ICASSP 2020} proposed the joint design of transmit precoding and detection rules using an autoencoder-based scheme, in which an end-to-end communication system is modeled by a DL network.
However, since they did not consider the typical modulations that can be easily realized by hardware, the resulting decision rules are too complicated to be implemented in practical systems.

Motivated by these findings, in this letter we propose an end-to-end learning architecture that jointly optimizes the transmitter and receivers with adaptive modulation orders and flexible symbol detection rules, to maximize the communication QoS in terms of symbol error rate (SER) under the constraints of power budget and total communication rate.
The considered MU-MISO system is modeled as a neural network that consists of a modulation order prediction (MOP-NN) module and a symbol-level precoding and detection (SLPD-NN) module.
Simulations illustrate the notable performance improvement brought by the proposed approach.

\section{System Model and Problem Formulation}

We consider a downlink MU-MISO system, where a BS equipped with $N_\text{t}$ antennas simultaneously serves $K$ single-antenna users.
Specifically, at the $l$-th time slot, the BS transfers one of $2^{M_k}$ messages to the $k$-th user, named $m_{k,l}$, $m_{k,l}\in\{1,\ldots, 2^{M_k}\}$, $M_k \in \{1, 2, 3, \ldots, B\}$, and $B$ is the allowed highest modulation order.
Considering that phase-shift keying (PSK) modulations are easier to implement in hardware and can exploit more constructive interference, we assume typical PSK modulations in this initial work.
Specially, the transmitted message $m_{k,l}$ is modulated as $s_{k,l} = \exp\{{\jmath2\pi m_{k,l}/M_k}\},~\forall k, l$.
Based on these assumptions, we investigate to adaptively optimize the modulation orders $\{M_1, \ldots, M_K\}$ in order to achieve better communication QoS.

For simplicity, we denote the modulated symbol vector at the $l$-th time slot as $\mathbf{s}_l\triangleq[s_{1,l}, \ldots, s_{K,l}]^T$.
According to the principle of SLP, the corresponding transmitted signal  $\mathbf{x}_l\in\mathbb{C}^{N_\text{t}}$ should be carefully designed to transfer $\mathbf{s}_l$.
This non-linear mapping can be expressed as $\mathbf{x}_l = \mathcal{P}(\mathbf{H},\mathbf{s}_l)$, where $\mathcal{P}(\cdot)$ represents the symbol-level precoding function, $\mathbf{H} \triangleq [\mathbf{h}_1, \mathbf{h}_2, \ldots, \mathbf{h}_K]$, and $\mathbf{h}_k\in\mathbb{C}^{N_\text{t}}$ denotes the channel vector between the BS and the $k$-th user.
Then, the received signal at the $k$-th user is written as
\be\label{eq:received signal}
r_{k,l} = \mathbf{h}_k^H\mathbf{x}_l + n_{k,l},
\ee
where $n_{k,l}\sim\mathcal{CN}(0,\sigma^2)$ is the additive white Gaussian noise (AWGN) of the $k$-th user.
After receiving the noise-corrupted signal $r_{k,l}$, the $k$-th user retrieves the transmitted message based on the available CSI $\mathbf{h}_k$ and (\ref{eq:received signal}) as $\widehat{m}_{k,l} = \mathcal{D}_k(\mathbf{h}_k, r_{k,l})$, where $\mathcal{D}_k(\cdot)$ represents the detection function for the $k$-th user.

We note that conventional SLP approaches assume fixed modulation orders $M_k$ and detection rules $\mathcal{D}_k,~\forall k$, under different CSIs.
Albeit this assumption provides a mathematically tractable model for optimizations, the flexibility of SLP is greatly limited considering that large variations in the channels may occur.
In such cases, intelligently adjusting the modulation orders as well as the detection rules will facilitate to achieve better performance.
Particularly, the SER of the $k$-th user can be written as
\be
\text{SER}_k = \mathbb{E}\{\text{Pr}(m_{k,l}\neq \widehat{m}_{k,l})\},
\ee
where $\text{Pr}(m_{k,l}\neq \widehat{m}_{k,l})$ represents the probability that the transmitted message $m_{k,l}$ is wrongly decoded as $\widehat{m}_{k,l}$.
Based on the above discussions, in this letter we aim to jointly optimize the modulation orders $M_k,~\forall k$, the precoding signal $\mathbf{x}_l$, and the detection rules $\mathcal{D}_k,~\forall k$, to minimize the average SER as well as to satisfy the total communication rate requirement $R$ and the transmit power budget $P$.
Thus, the optimization problem can be formulated as
\begin{subequations}
\label{eq:original problem}
\begin{align}
&\underset{\mathbf{x}_l, \mathcal{D}_k, M_k, \forall k}{\min}\quad\frac{1}{K}\sum_{k=1}^K\text{SER}_k \\
&\hspace{0.6 cm}\text{s.t.}\hspace{1 cm}\mathbf{x}_l = \mathcal{P}(\mathbf{H},\mathbf{s}_l),\\
&\hspace{2 cm}\widehat{m}_{k,l} = \mathcal{D}_k(\mathbf{h}_k,r_{k,l}),\\
&\hspace{2 cm}\sum_{k=1}^KM_k\geq R, \\
&\hspace{2 cm}\|\mathbf{x}_l\|^2 \leq P.
\end{align}
\end{subequations}
It is obvious that problem (\ref{eq:original problem}) cannot be directly solved by traditional optimization algorithms, and even heuristic algorithms would have excessive computational complexities.
In the next section, we propose an end-to-end neural network, named adaptive modulation symbol-level precoding and detection neural network (AMPD-NN), for the joint design of symbol-level precoding and detection with adaptive modulation orders.

\begin{figure}[t]
\centering
\includegraphics[width = \linewidth]{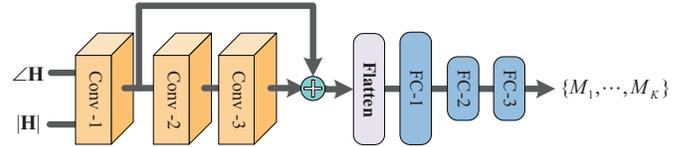}\vspace{0.1 cm}
\caption{Structure of MOP-NN.}
\label{fig:MOP-NN}
\end{figure}

\begin{figure*}[t]
\centering
\includegraphics[width = 0.85\linewidth]{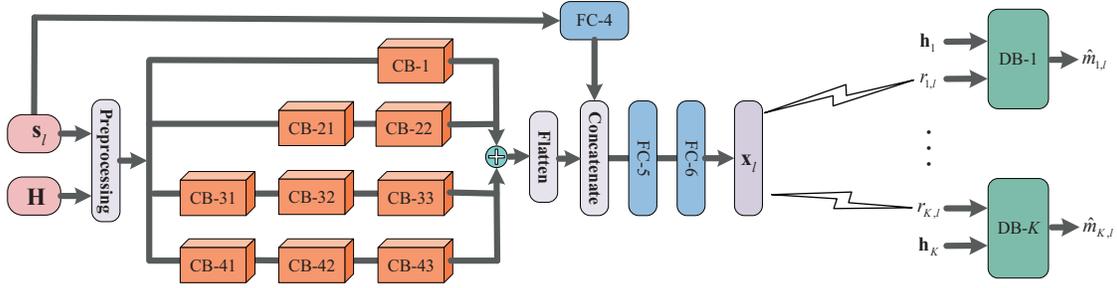}\vspace{0.1 cm}
\caption{Structure of SLPD-NN.}
\label{fig:SLPD-NN}
\end{figure*}

\section{An End-to-End Neural Network for Symbol-Level Precoding and Detection with Adaptive Modulation}

\subsection{Network Structure}

The proposed AMPD-NN consists of an MOP-NN module and an SLPD-NN module.
In particular, MOP-NN is designed to predict the modulation orders, i.e., perform a multi-classification task to determine the values of $M_k, \forall k$.
For modulated symbols that may have different modulation orders, the SLPD-NN intelligently optimizes symbol-level precoding and detection rules to realize end-to-end communication.
Details about the network structures of these two modules are presented below.

The compact structure of MOP-NN is presented in Fig. \ref{fig:MOP-NN}, of which the input is the amplitude and phase information of $\mathbf{H}$ and the output is the modulation orders $\{M_1, \ldots, M_K\}$.
The final fully connected (FC) layer adopts a softmax activation function to obtain the probabilities of different categories.
In specific, the categories are composed of all possible modulation order combinations, which satisfy the constraints of the total communication rate $R$ and the highest modulation order $B$. These categories are sequentially numbered and stored in a priori. After obtaining the index of the category with the highest probability, the modulation orders $\{M_1, \ldots, M_K\}$ can be accordingly selected.
In addition, a batch normalization layer is added after each convolutional layer (CL), FC-2, and FC-3, and a dropout layer is added after the activation function of FC-1 layer, to prevent overfitting.

The structure of SLPD-NN is illustrated in Fig. \ref{fig:SLPD-NN} at the top of the next page.
The modulated symbol vector $\mathbf{s}_l$ and the CSI $\mathbf{H}$ are passed through the neural network at the BS to generate the transmitted precoding signal $\mathbf{x}_l$, and then the receivers use the received signals and CSI to decode the transmitted messages.
Considering that the computing ability at the transmitter is usually more powerful than the receivers, we deploy complicated CNNs at the BS and several FC layers at the receivers.
Specifically, a parallel CNN architecture with different numbers of convolutional blocks (CBs) is utilized to extract useful information from the input with various modulation orders.
Four parallel lines with convolutional kernels of different sizes enable the network to extract features in multi-scale regions.
Each CB consists of four CLs, where the first CL has $8\times (1,1)$ filters and the other three layers have $8\times (1,d)$ filters (the parameter $d$ has different values for different CBs as shown in Table I), each with a ``SAME'' padding and $(1,1)$ stride, followed by a LeakyReLU activation function and a batch normalization layer.
In addition to the fusion of $\mathbf{s}_l$ and $\mathbf{H}$, the symbol information $\mathbf{s}_l$ is also fed into this neural network though FC-4, and its output is concatenated with the vectorized features extracted by the parallel CNN architecture.
This is similar to a residual connection through which information from the shallow layers flows into the deeper layers, thus enhancing the impact of the input data.
Then, the obtained information is regressed through two FC layers to output the corresponding precoding, which is normalized to meet the power constraint before being forwarded to the transmit antennas.
At the user side, the same decode block (DB) composed of four FC layers (denoted as FC-D1/D2/D3/D4) is applied to decode the desired messages.
The input of each DB is the received noise-corrupted signal and the corresponding CSI, the output is the probability distribution over all possible messages.
We pick the index with the highest probability as the decoded message.
The parameters and settings of the proposed AMPD-NN are summarized in Table I.

\begin{table}[!t]
\centering
\caption{\label{tab:parameters}Parameters and settings of AMPD-NN.}
\vspace{-0.1 cm}
\begin{center}\begin{small}
\begin{tabular}{ c  c  c  }
   \hline
    Layer         & Activation function    & Setting     \\
   \hline
   Conv-1/2/3     & ReLU                   & $4\times(1,1)$  \\
   FC-1           & ReLU                   & 32 \\
   FC-2           & ReLU                   & $N_\text{m}$  \\
   FC-3           & Softmax                & $N_\text{m}$  \\
   CB-1/22/33/43 & LeakyReLU              & $d = 1$ \\
   CB-21/32/42    & LeakyReLU              & $d=3$ \\
   CB-31          &LeakyReLU               &$d = 5$  \\
   CB-41          &LeakyReLU               &$d = 7$  \\
   FC-4           &LeakyReLU               &32            \\
   FC-5           &LeakyReLU               &256           \\
   FC-6           &none                    &$2N_\text{t}$   \\
   FC-D1                 &LeakyReLU             &128           \\
   FC-D2                 &LeakyReLU             &64            \\
   FC-D3                 &LeakyReLU             &32            \\
   FC-D4                 &Softmax               &$B$   \\
   \hline
\end{tabular}\end{small}
\end{center}
\end{table}

\subsection{Input Pre-processing}

To capture more features of the data, proper pre-processing should be performed before feeding the raw data into the neural network.
Since the existing deep learning libraries cannot handle complex data and the considered PSK modulations rely on the phase information of signals, the complex-valued CSI is separated into phase information $\angle\mathbf{H}$ and amplitude information $|\mathbf{H}|$ as the input to the MOP-NN.
For the SLPD-NN, the phase information of the transmitted symbols (i.e., $\angle\mathbf{s}_l$) is the input fed into the neural network through FC-4.
In addition, $\angle\mathbf{s}_l$ and $\mathbf{H}$ are pre-processed and forwarded to the CLs as another input of SLPD-NN.
The aim is to avoid the neural network overly focusing on the specific phase information and ignoring the overall information.
Particularly, by properly replicating $\angle\mathbf{s}_l$ and then adding it to $\angle{\mathbf{H}}$, the phase information of the desired symbol is added to the corresponding channel as
\be
\widetilde{\mathbf{H}} = \angle\mathbf{H} + \mathbf{1}\angle\mathbf{s}^T_l.
\ee
The input enables the neural network to learn a more flexible decision rule, which may be a rotated version of the input.

\subsection{Loss Functions}

The goal of the proposed end-to-end neural network is to jointly optimize transmit precoding and receive detection to successfully decode the transmitted messages at the users.
This task can be seen as a multi-classification problem, which is generally solved by stochastic gradient descent (SGD) algorithm associated with a categorical cross-entropy loss function.
Thus, the loss function of the SLPD-NN can be written as the cross-entropy between the one-hot representations of the transmitted messages and the probability vectors generated at the users.
The one-hot encoding of $m_{k,l}$ is usually represented as $\mathbf{t}_{k,l}\in\{0,1\}^{2^{M_k}}$, where the $m_{k,l}$-th element of $\mathbf{t}_{k,l}$ is one and the others are zeros.
Note that the multi-classification task is actually different for each user, since the modulation orders $M_k,~\forall k$, are adaptively optimized and the corresponding labels $\mathbf{t}_{k,l},~\forall k$, are of different sizes.
Instead of training different layers for all possible modulation orders, we recast the multi-classification tasks for different modulation orders as a multi-classification task for the highest modulation order.
This allows the users to employ the same detection architecture, thus significantly reducing the complexity of the network and training.
In particular, the one-hot encoding of $m_{k,l}$ is re-formulated as $\widetilde{\mathbf{t}}_{k,l}\in\{0,1\}^{2^B}$ with some zeros appended at the end of $\mathbf{t}_{k,l}$.
Accordingly, the output probability vector is denoted as $\widetilde{\mathbf{p}}_{k,l}$, which satisfies $\sum_{i=1}^{2^{B}}\widetilde{\mathbf{p}}_{k,l}(i) = 1$ and $\widetilde{\mathbf{p}}_{k,l}(i) \geq 0, \forall i$.
Thus, the cross-entropy loss function of SLPD-NN is written as
\be
{L}_\text{SLPD} = \frac{1}{N}\sum_{n=1}^NL_n = -\frac{1}{N}\sum_{n=1}^N\sum_{i=1}^{2^{B}}\widetilde{\mathbf{t}}_{k,l}(i)\log(\widetilde{\mathbf{p}}_{k,l}(i)),
\ee
where $N$ is the mini-batch size.
Similarly, the loss function of MOP-NN is formulated as
\be
{L}_\text{MOP} = -\frac{1}{N}\sum_{n=1}^N\sum_{i=1}^{N_\text{m}}{\mathbf{t}}_{\text{MOP}}(i)\log(\mathbf{p}_{\text{MOP}}(i)),
\ee
where $\mathbf{t}_\text{MOP}\in\{0,1\}^{N_\text{m}}$ is the one-hot representation of the input label of $\{M_1, \ldots, M_K\}$, $N_\text{m}$ is the number of possible modulation order combinations under the constraints of the total communication rate and the highest modulation order, and $\mathbf{p}_{\text{MOP}}$ is the output probability vector.

\subsection{Training Strategy}

A three-stage training strategy is developed to train the proposed AMPD-NN.
Considering that the tasks of the SLPD-NN are similar for different modulation orders, we propose to utilize the weights obtained in the pre-training process to initialize the neural network.
This so-called transfer learning strategy can accelerate the training process and facilitate convergence to a better result.
In stage I, we pre-train the SLPD-NN with given fixed modulation orders, and then store the obtained weights for subsequent use.
In stage II, the SLPD-NN is trained with randomly generated modulation orders using the initial weights obtained in stage I.
These modulation order combinations should satisfy the constraints of the highest modulation order $B$ and the total communication rate $R$.
After that, the modulation order combination corresponding to the lowest cross-entropy achieved in each channel realization is set as the label for supervising the training of MOP-NN.
Finally, the MOP-NN is trained with the labels obtained in stage II.

The details of offline training are described as follows.
We first generate $1.2e^{5}$ channel realizations, of which $1e^{5}$ for training, $1e^{4}$ for testing, and $1e^{4}$ for validation.
The typical Adam optimizer is utilized and the mini-batch size is $1000$.
The learning rate is initialized as $0.001$ and then decayed by $0.1$ per $50$ epoches.
In each epoch, each channel sample is trained by $5$ times with randomly generated modulation orders and transmitted symbols.
Similar to \cite{Sohrabi ICASSP 2020}, random noise is added during training so that the trained network can adapt to different transmit signal-to-noise ratios (SNRs), which are defined as $\Gamma \triangleq 10\log_{10}{\frac{P}{\sigma^2}}$ and uniformly distributed within a reasonable range of $[0,20]$ dB.
In stage I, the SLPD-NN is trained by $100$ epoches with fixed quadrature-PSK (QPSK) modulations.
Then, the SLPD-NN is trained with all possible modulation order combinations under the constraints of the highest modulation order and the total communication rate for another $100$ epoches, during which the labels are generated.
In stage III, the MOP-NN is trained for $50$ epoches.

\subsection{Practical Deployment}

After finishing the offline training process described in the previous subsections, the proposed AMPD-NN can be deployed.
In particular, the trained MOP-NN is deployed at the BS to predict the modulation orders for given CSI.
The trained SLPD-NN is separated into two parts and deployed at the BS and the users, in which partial neural network at the BS generates transmitted signals with given modulated symbols and CSI, and partial neural network at the users decodes the desired messages with CSI and received signals.

\section{Simulation Results}

\begin{figure}[t]
\centering
\includegraphics[width = 0.95\linewidth]{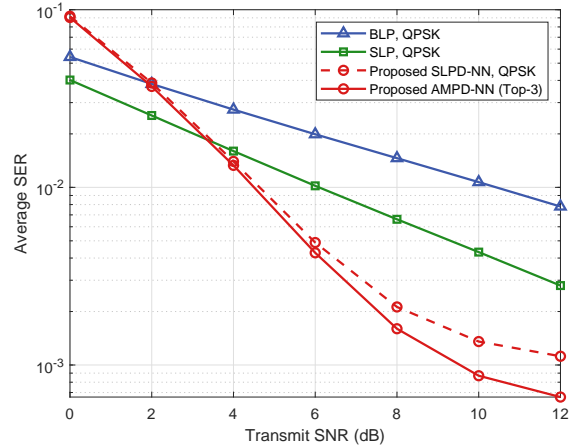}
\caption{Average SER versus transmit SNR.}
\label{fig:SER}
\end{figure}

In this section, we present simulations to validate the effectiveness of the proposed end-to-end learning approach for symbol-level precoding and detection with adaptive modulation.
We assume that the BS equipped with $N_\text{t} = 128$ antennas serves $K=4$ single-antenna users.
The highest modulation order is set as $B=3$, i.e., binary-PSK (BPSK), QPSK, and $8$-PSK can be adaptively selected.
The requirement of the total communication rate is set as $R = 8$ bits per channel use.
We assume a limited-scattering environment as that in \cite{Sohrabi ICASSP 2020}, where each user has one scatter path.
The angle of arrival (AoA) of the $k$-th user is uniformly distributed in $\theta_k \sim [\phi_k-10^\circ,\phi_k+10^\circ]$, and $\{\phi_1, \phi_2, \phi_3, \phi_4\} = \{-30^\circ, -15^\circ, 15^\circ, 30^\circ\}$.
The simulations are implemented by using TensorFlow $2.1.0$, Keras $2.3.1$, and Python $3.7.7$ on a computer with NVIDIA TITAN Xp GPU and Intel(R) Core(TM) i9-7900X CPU.

We first illustrate the average SER versus transmit SNR in Fig. \ref{fig:SER}.
Following the routine of evaluating multi-classification performance, we use the Top-3 options of MOP-NN output and select the best result from them named as ``Proposed AMPD-NN (Top-3)''.
For comparisons, we also include the BLP, the conventional optimization based SLP, and the proposed end-to-end SLPD-NN approaches with fixed QPSK modulations (denoted as ``BLP, QPSK'', ``SLP, QPSK'', ``Proposed SLPD-NN, QPSK'', respectively).
It is clear that the average SER of the proposed end-to-end learning based approaches decrease faster than their counterparts and achieve much better performance during a reasonable transmit SNR range.
For a 10dB transmit SNR, for example, the average SER of the proposed approach is below $1e^{-3}$, while that of the conventional SLP approach with QPSK modulations is about $5e^{-3}$ and the BLP approach is about $1e^{-2}$.
In addition, the proposed approach with adaptive modulation orders shows increasingly pronounced performance improvements with increasing transmit SNR compared to the scheme with fixed modulation orders.
These results validate the effectiveness of the proposed end-to-end learning approach in joint design of the modulation order, the transmit symbol-level precoding and the receive detection.

Next, the modulation order prediction accuracy for the training and testing sets and the average SER for the validation set are presented in Table II, where the prediction results for Top-1, Top-2 and Top-3 are shown and the transmit SNR is set as 15dB.
It is noted that the scheme using Top-3 achieves the best performance with minimal complexity sacrifice, which makes it a good choice in practice.
Moreover, we observe that the prediction accuracy can reach $96.79\%$ when 50 different modulation order combinations exist in the settings of this paper, which verifies the effectiveness of the proposed adaptive modulation order design.

\begin{table}[t]
\centering\begin{small}
\caption{\label{tab:simulation}Prediction accuracy and average SER.}
\vspace{-0.1 cm}
\begin{center}
\begin{tabular}{ c  c  c  c  }
   \hline
            & training accuracy     & testing accuracy      & average SER \\
   \hline
   Top-1    & $0.8226$                 &$0.8049$                 &$1.10e^{-3}$  \\

   Top-2    &$0.9380$                 &$0.9304$                 &$8.28e^{-4}$  \\

   Top-3    &$0.9685$                 &$0.9679$                 &$6.60e^{-4}$  \\
   \hline
\end{tabular}
\end{center}\end{small}
\vspace{-0.1 cm}
\end{table}

\begin{figure}[t]
\centering
\subfigure[Proposed SLPD-NN with QPSK modulations.]{
\begin{minipage}{\linewidth}
\centering
\includegraphics[width = \linewidth]{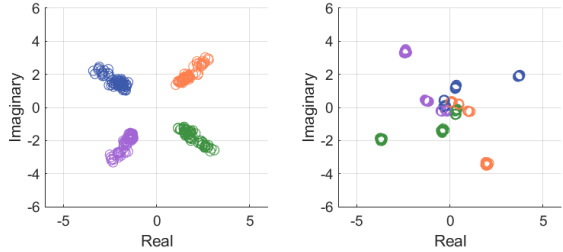}\vspace{0.2 cm}
\end{minipage}}
\vspace{0.1 cm}
\subfigure[Proposed AMPD-NN with adaptive modulations.]{
\begin{minipage}{\linewidth}
\centering
\includegraphics[width = \linewidth]{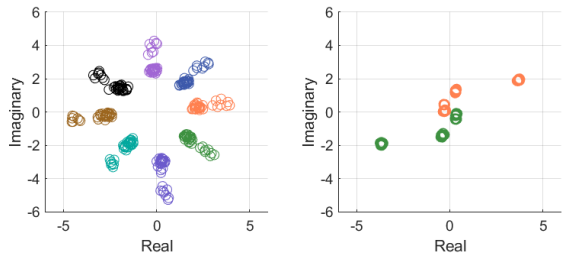}\vspace{0.2 cm}
\end{minipage}}
\caption{Received noise-free signals (left: user-1, right: user-2).}
\label{fig:constellations}
\end{figure}

Finally, the received noise-free signals of user-1 and user-2 are shown in Fig. \ref{fig:constellations} to visualize the advantages of the proposed approach.
For the same channel realization, the proposed SLPD-NN approach with fixed QPSK modulations (i.e., $M_1 = M_2 = M_3 = M_4 = 2$) is presented in Fig. \ref{fig:constellations}(a), and the proposed AMPD-NN approach is illustrated in Fig. \ref{fig:constellations}(b) where $M_1 = 3$, $M_2 = 1$, $M_3 = 2$, and $M_4 = 2$.
It can be observed in Fig. \ref{fig:constellations}(a) that, the decision rules are not the same for different users although the same QPSK modulation is chosen.
This phenomena indicates that the end-to-end learning approach allows more flexibility to design different detection rules at the users to facilitate transmissions.
In addition, we can see that the channel of user-1 is much better than that of user-2, since the received signals of the first user are much more scattered.
In such cases, the scheme with the same fixed modulations cannot guarantee satisfactory fairness, while the proposed adaptive modulation based approach will achieve much better performance.
Particularly, a lower modulation order (i.e., BPSK) is chosen for user-2 to improve the SER performance and a higher modulation order (i.e., 8-PSK) for user-1 to ensure the total communication rate performance.
As shown in Fig. \ref{fig:constellations}(b), we clearly see that the received signals of user-2 is much more scattered than its counterpart in Fig. \ref{fig:constellations}(a).
These results demonstrate the advantages of the proposed end-to-end learning based adaptive modulation approach.

\section{Conclusion}

We proposed an end-to-end learning approach to jointly optimize the modulation order, the transmit symbol-level precoding and the receive detection.
We established a neural network consisting of the modulation order prediction module and the symbol-level precoding and detection module, and trained it using cross-entropy loss functions and the proposed three-stage training strategy.
Simulations validated the flexibility and advantages of the proposed end-to-end learning approach.


\begin{thebibliography}{99}

\bibitem{Masouros TSP 2015} C. Masouros and G. Zheng, ``Exploiting known interference as green signal power for downlink beamforming optimization,'' \textit{IEEE Trans. Signal Process.}, vol. 63, no. 14, pp. 3628-3640, Jul. 2015.
\bibitem{Alodeh ICST 2018} M. Alodeh, \textit{et al.}, ``Symbol-level and multicast precoding for multiuser multiantenna downlink: A state-of-art, classification, and challenges,'' \textit{IEEE Commun. Surveys Tut.}, vol. 20, no. 3, pp. 1733-1757, May 2018.

\bibitem{Li ICST 2020} A. Li, \textit{et al.}, ``A tutorial on interference exploitation via symbol-level precoding: Overview, state-of-the-art and future directions,'' \textit{IEEE Commun. Surveys Tut.}, vol. 22, no. 2, pp. 796-839, 2nd Quart. 2020.

\bibitem{AL ITWC 2018} A. Li and C. Masouros, ``Interference exploitation precoding made practical: Optimal closed-form solutions for PSK modulations,'' \textit{IEEE Trans. Wireless Commun.}, vol. 17, no. 11, pp. 7661-7676, Nov. 2018.

\bibitem{Xiao TCOM 2022} Z. Xiao, R. Liu, M. Li, Y. Liu, and Q. Liu, ``Low-complexity designs of symbol-level precoding for MU-MISO systems,'' \textit{IEEE Trans. Commun.}, vol. 70, no. 7, pp. 4624-4639, Jul. 2022.

\bibitem{Shea TCCN 2017} T. O'Shea and J. Hoydis, ``An introduction to deep learning for the physical layer,'' \textit{IEEE Trans. Cogn. Commun. Netw.}, vol. 3, no. 4, pp. 563-575, Dec. 2017.

\bibitem{Dorner JSTSP 2018} S. D\"{o}rner, S. Cammerer, J. Hoydis, and S. Brink, ``Deep learning based communication over the air,'' \textit{IEEE J. Sel. Topics Signal Process.}, vol. 12, no. 1, pp. 132-143, Feb. 2018.

\bibitem{Zhang ICST 2019} C. Zhang, P. Patras, and H. Haddadi, ``Deep learning in mobile and wireless networking: A survey,'' \textit{IEEE Commun. Surveys Tut.}, vol. 21, no. 3, pp. 2224-2287, 3rd Quart. 2019.

\bibitem{Lei CL 2021} Z. Lei, X. Liao, Z. Gao, and A. Li, ``CI-NN: A model-driven deep learning-based constructive interference precoding scheme,'' \textit{IEEE Commun. Lett.}, vol. 25, no. 6, pp. 1896-1900, Jun. 2021.

\bibitem{Zhu TVT 2021} Z. Bo, R. Liu, M. Li, and Q. Liu, ``Deep learning based efficient symbol-level precoding design for MU-MISO systems,'' \textit{IEEE Trans. Veh. Technol.}, vol. 70, no. 8, pp. 8309-8313, Aug. 2021.

\bibitem{Mohammad GLOBECOM 2021} A. Mohammad, C. Masouros, and Y. Andreopoulos, ``An unsupervised learning-based approach for symbol-level precoding,'' in \textit{Proc. IEEE Global Commun. Conf. (GLOBECOM)}, Madrid, Spain, Dec. 2021.

\bibitem{Sohrabi ICASSP 2020} F. Sohrabi, H. V. Cheng, and W. Yu, ``Robust symbol-level precoding via autoencoder-based deep learning,'' in \textit{Proc. IEEE Int. Conf. Acoustics Speech Signal Process. (ICASSP)}, Barcelona, Spain, May 2020.

\end{thebibliography}
\end{document}